\documentclass[epjCONF]{svjour}
\usepackage{graphics}
\usepackage[varg]{txfonts} 
\usepackage[latin1]{inputenc}
\session-title{New Technologies for Probing the Diversity of Brown Dwarfs and Exoplanets}
\begin{document}
\title{CoRoT's first seven planets: An Overview
\thanks{The CoRoT space mission has been developed and is operated by CNES with the contribution of Austria, Belgium, Brazil, ESA, Germany and Spain.}}
\author{R. Dvorak \inst{1}, J. Schneider \inst{2}, H. Lammer \inst{3}, P. Barge \inst{4}, G. Wuchterl \inst{5} and the CoRoT team}
%

\institute{Institute for Astronomy, T\"urkenschanzstrasse 17, A-1180 Vienna, Austria \and
LUTH, Observatoire de Paris-Meudon, 5 place J.Jansen, F-92195 Meudon, Paris, France \and
Space Research Institute, Austrian Academy of Sciences, Schmiedlstrasse 6, A-8042 Graz, Austria \and
Laboratoire d'Astrophysique de Marseille, Technopole de Marseille-Etoile, F-13388 Marseille Cedex 13, France \and
Th\"uringer Landessternwarte Tautenburg, Sternwarte 5, D-07778, Tautenburg, Germany}

\abstract{
The up to 150 day uninterrupted high-precision photometry of about
100000 stars -- provided so far by the exoplanet channel of the CoRoT
space telescope -- gave a new perspective on the planet population of our
galactic neighbourhood. The seven planets with very accurate parameters widen the
range of known planet properties in almost any respect. Giant planets
have been detected at low metallicity, rapidly rotating and active,
spotted stars. CoRoT-3 populated the brown dwarf desert and closed
the gap of measured physical properties between standard giant planets
and very low mass stars. CoRoT extended the known range of planet masses
down-to 5 Earth masses and up to 21 Jupiter masses, the radii to less
than 2 Earth radii and up to the most inflated hot Jupiter found so
far, and the periods of planets discovered by transits to 9 days.
Two CoRoT planets have host stars with the lowest content
of heavy elements known to show a transit hinting towards a different
planet-host-star-metallicity relation then the one found by
radial-velocity search programs. Finally the properties of the CoRoT-7b
prove that terrestrial planets with a density close to Earth exist
outside the Solar System. The detection of the secondary transit of
CoRoT-1 at the $10^{-5}$-level and the very clear detection of the 1.7
Earth radii of CoRoT-7b at $3.5 \; 10^{-4}$ relative flux are promising evidence
of CoRoT being able to detect even smaller, Earth sized planets.
} 

\maketitle

\section{Introduction}
\label{intro}
The space mission CoRoT is a joint adventure of different European
countries, namely CNES, the French space agency as leading organisation
and other ones in Austria, Brazil, Belgium, Germany and Spain and also the
European Space Agency ESA.
The original goal was to
observe variable stars but -- with the discovery of the first exoplanet
in the early 90's -- the search for exoplanets was
included in the research program (see Schneider and Chevreton (\cite{Sch90}) and
turned out to be also
very important. Already in the name of the mission CoRoT both research tasks are
mentioned {\bf Co}nvection, {\bf Ro}\-tation and planetary {\bf T}ransits.
The satellite was successfully launched on December, 27th 2006 from Baikonour,
observations started in February 2007 and first results concerning extrasolar
planet transits respectively an overview of the CoRoT mission were published
by Barge et al. \cite{Bar08a}. The life-time was scheduled for 3 years of
observations and recently the extension for another three years was approved.
Some of the results obtained with measurements of the CoRoT mission were
 published in a whole volume of the
journal Astronomy and Astrophysics this year (Baglin et al. \cite{Bag09}).

Because there exist already several introductionary papers describing the
mission and the instrument in detail (like the one cited above by Barge et al.)
we just mention the main characteristics of the
satellite: The 4~m long and 630~kg massive satellite has a mean diameter of 2~m.
The accuracy  of the pointing is 0.5 arcsec with a capacity of the telemetry
of 1.5 Gbit/day. Three systems are operating on the satellite:

\begin{enumerate}
\item {\bf CoRoTel} is
an afocal telescope with 2 parabolic mirrors and aperture of 27 cm a
cylindric baffle 2 m long,
\item {\bf CoRoTcam} is a wide field
camera consisting of a
dioptric objective of 6 lenses where the focal unit is equip\-ped with 4
frame-transfer-CCD 2048x4096. Two of the four CCDs are for the exoplanet program and
two are for the seismology program.
\item {\bf CoRoTcase} is hosting the
electronics and the software managing the aperture photometry processing.
\end{enumerate}

For the exoplanet CCD the total number of stars is 12000 and the magnitudes
are between 11 and 16; the flux is measured every 512 s consisting of 16
individual exposures of 32 s. A bi-prism in the focal block dedicated to
exoplanets allows to
get chromatic information (red, green and blue)
for brighter stars, but for faint stars only the standard one band (white)
photometry is performed. The polar orbit of the satellite allows to observe
150 days in the
long run, and between 20 and 30 days in the short run, both in the
direction of the center respectively anticenter of the milky way, which
makes two reversal manoeuvres necessary per year.

Radial velocity measurements to confirm planet candidates were usually performed with the following
instruments:

\begin{enumerate}
\item HARPS spectrograph in La Silla  (3.6 m telescope)
\item SOPHIE spectrograph in the Observatoire de Haute\\
 Provence (1.93 m telescope)
\item CORALIE spectrograph on the 1.2 swiss telescope in La Silla
\item Coud\'e echelle spectrograph from the 2 m telescope Tautenburg (TLS)
\item UVES and FLAMES spectrographs in Chile
\item Spectrographs at McDonald observatory
\end{enumerate}

\section{CoRoT-1b: The first planet discovered by CoRoT}

\begin{figure}
\begin{center}
\resizebox{1.00\columnwidth}{!}{
  \includegraphics{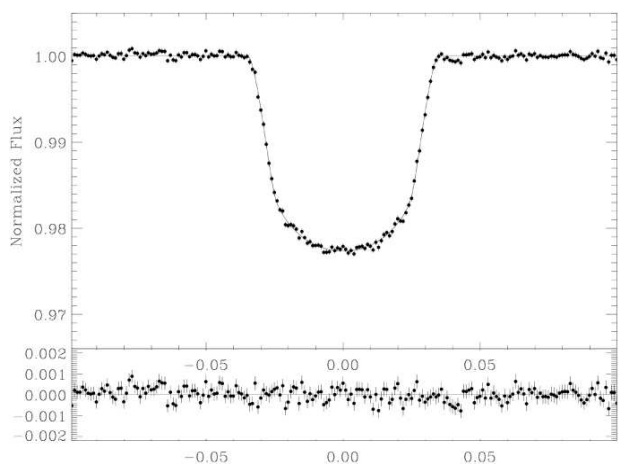} }
\end{center}
\caption{Normalised and phase-folded lightcurve of the best 34 transits of
  CoRoT-1b (\cite{Bar08b} Fig.1).}
\label{1-1}       
\end{figure}

\begin{figure}
\begin{center}
\resizebox{1.00\columnwidth}{!}{
  \includegraphics{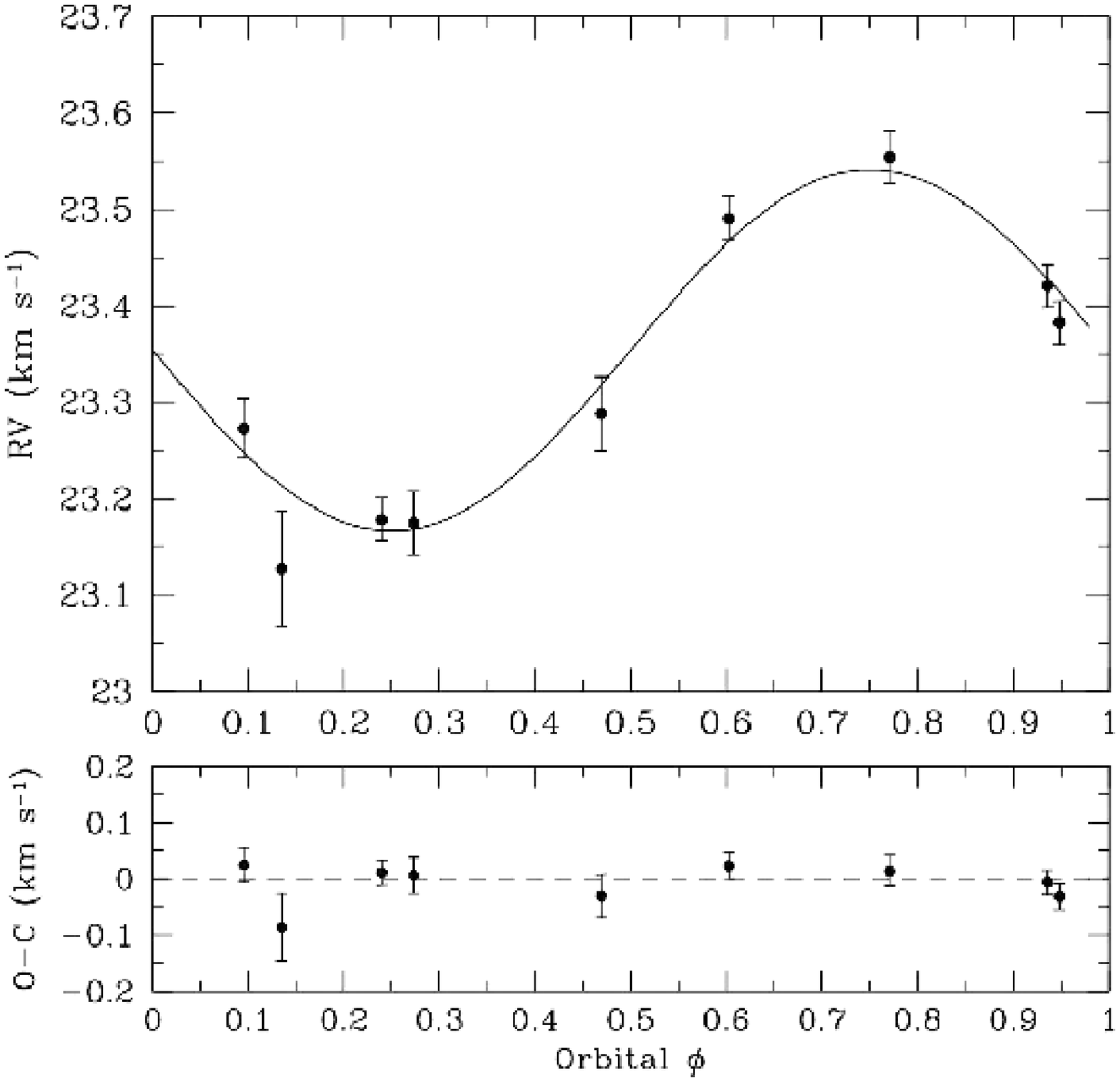} }
\end{center}
\caption{RV measurements of CoRoT-1b with SOPHIE (\cite{Bar08b} Fig.2).}
\label{1-2}       
\end{figure}

\begin{figure}
\begin{center}
\resizebox{1.00\columnwidth}{!}{
  \includegraphics{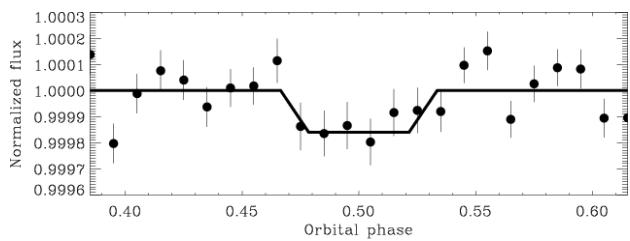} }
\end{center}
\caption{The secondary eclipse of CoRoT-1b  (\cite{Alo09}, Fig.5).}
\label{1-3}       
\end{figure}

Already the first exoplanet discovered by CoRoT measurements was a surprise
concerning its nature (Barge et al. \cite{Bar08b}): it turned out to have a very small density ($0.38\, g\, cm^{-3}$)\footnote{In the text all the parameters are given without
the error bars. The complete values with the estimated errors are given in the appendix} compared to all other planets found up to now. The short
period of 1.5 days -- the light curve is shown in Fig. \ref{1-1} -- gave a
very well determined
period because 34 successive
transits could be used for its computation. Additional RV observation
(Fig. \ref{1-2}) were
undertaken from OHP with the SOPHIE spectrograph for 9 different positions of
the exoplanet and confirmed the planetary nature of the transit of a planet with
a radius of
1.49 $R_{Jup}$ and a mass  of 1.03 $M_{Jup}$ in front of a G0V star. Later
photometric observation revealed the secondary eclipse, when the planet
disappears behind the disc of the star which was discovered despite a very
shallow nature of the signal (Fig. \ref{1-3}).

\section{CoRoT-2b: a transiting planet around an active G star}

\begin{figure}
\begin{center}
\resizebox{1.00\columnwidth}{!}{
  \includegraphics{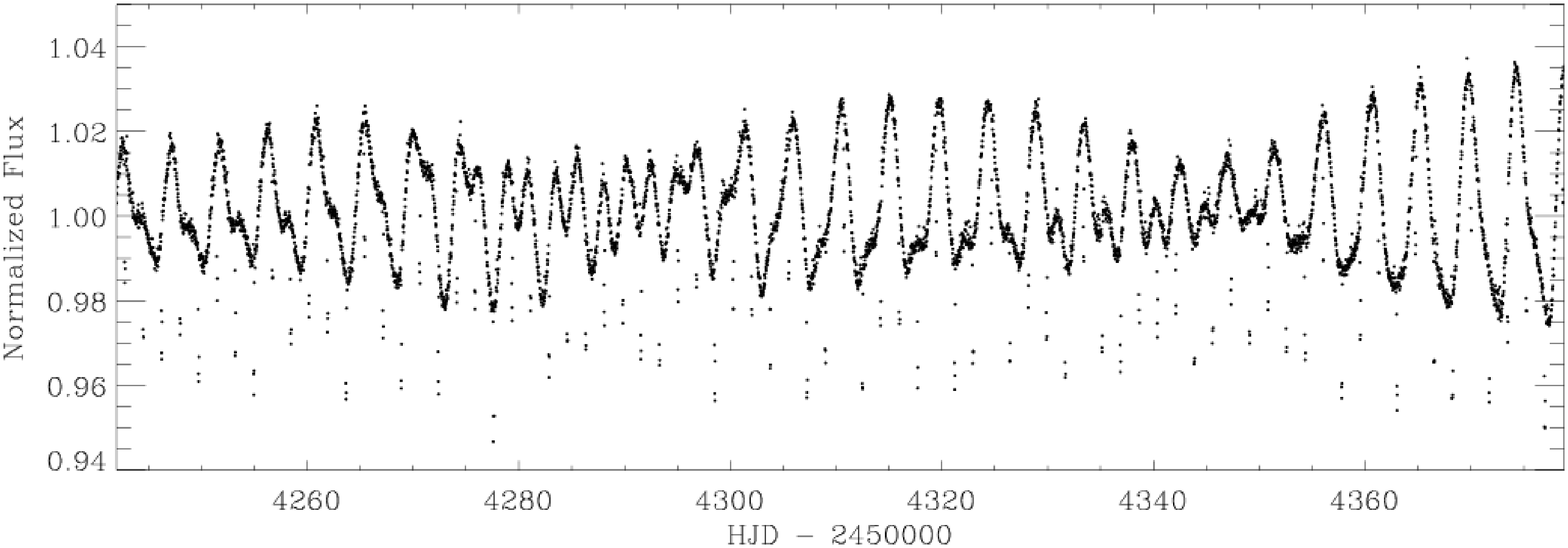} }
\end{center}
\caption{Normalized flux of CoRoT-2b using 78 transits (\cite{Alo08}, Fig.1).}
\label{2-1}       
\end{figure}

\begin{figure}
\begin{center}
\resizebox{1.00\columnwidth}{!}{
  \includegraphics{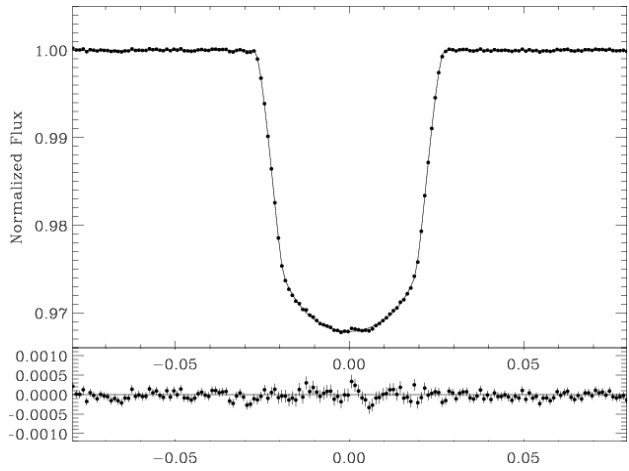} }
\end{center}
\caption{Normalized and phase folded light curve of 78 transits of CoRoT-2b (\cite{Alo08}, Fig.2).}
\label{2-2}       
\end{figure}

\begin{figure}
\begin{center}
\resizebox{1.00\columnwidth}{!}{
  \includegraphics{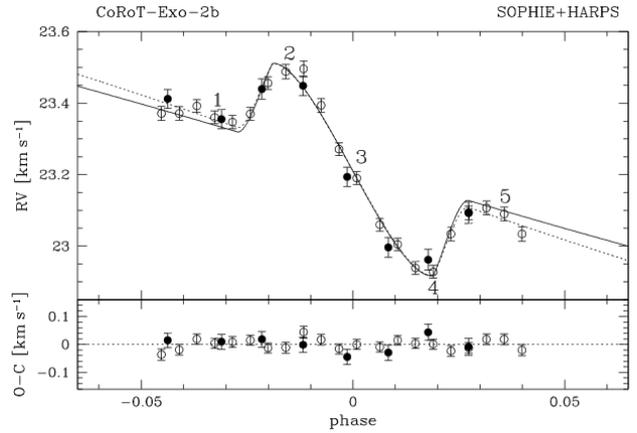} }
\end{center}
\caption{Phase-folded RV measurements of CoRoT-2b derived
  by SOPHIE and HARPS showing the Rossiter-McLaughlin effect during the transit (\cite{Bou08}, Fig.1).}
\label{2-4}       
\end{figure}

Out of 78 transits during the 150 days of observations of CoRoT-2b (Alonso et
al. \cite{Alo08}) (Fig. \ref{2-1}) the composite
light curve for the transit (Fig. \ref{2-2}) was derived after
removing the signal of the star's rotation and activity, which is due to large
spots on its surface. The respective periods were between 4.5 and 5 days and
showed flux variations of a few percents. This second confirmed planet of CoRoT
orbits a K0V star with a metallicity of [Fe/H]=0 in 1.74 days and has a mass
of 3.3 $M_{Jup}$ and a radius of 1.465 $R_{Jup}$. With the use of very precise
RV measurements with SOPHIE, HARPS and CORALIE it was possible to determine an inclination of
$\lambda=7.2^{\circ} $ between the planetary orbit and the equator of the
central star (\cite{Bou08}). The radial velocity anomaly during the transit of the planet
(Fig. \ref{2-4})--
the Rossiter-McLaughlin effect\footnote{
The rotation of a star leads to a line broadening in the spectra because
by the coming towards the observer and the moving away of different parts of
the disc, consequently a blueshift
respectively a redshift in the star's spectrum is observable. Now when the
planet hosting star is in transit, it hides part of the disc which causes the
observed mean redshift to vary and this redshift anomaly switches from
negative to positive (or vice versa) and of the form of this anomaly the
inclination of the planets' orbit with respect to the equator of the star can
be determined.} was measured up to now only for 5
transiting exoplanets. This new measurements
showed that even fainter stars ($m_v \le 12$) transited by a planet can be
used to determine this effect with telescopes of the 2~m class.

\section{CoRoT-3b: the first secure inhabitant of the brown dwarf desert}

\begin{figure}
\begin{center}
\resizebox{1.00\columnwidth}{!}{
  \includegraphics{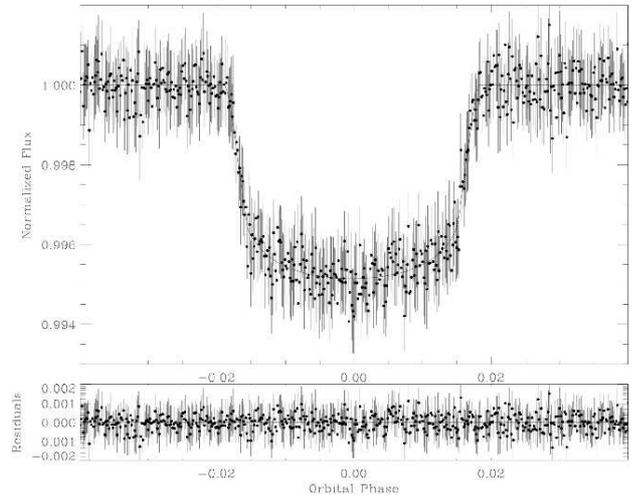} }
\end{center}
\caption{Light curve of CoRoT-3b (\cite{Del08}, Fig.2).}
\label{3-1}       
\end{figure}

\begin{figure}
\begin{center}
\resizebox{1.00\columnwidth}{!}{
  \includegraphics{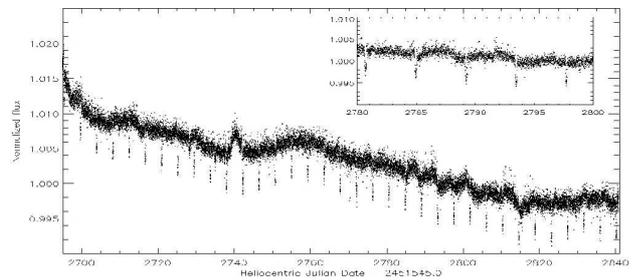} }
\end{center}
\caption{Light curve of CoRoT-3b for 152 days (\cite{Del08}, Fig.1).}
\label{3-2}       
\end{figure}

\begin{figure}
\begin{center}
\resizebox{1.00\columnwidth}{!}{
  \includegraphics{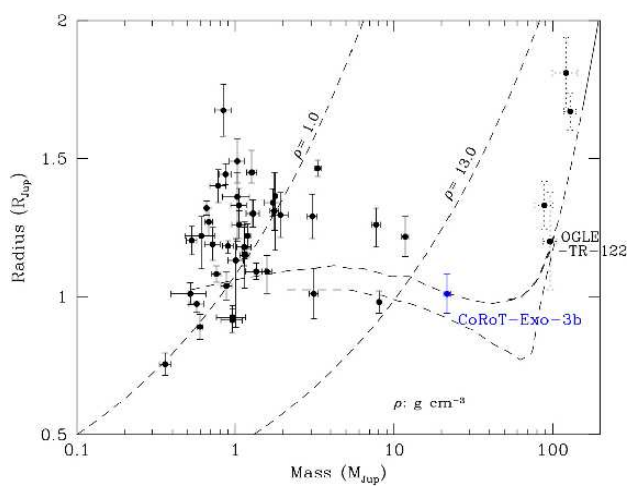} }
\end{center}
\caption{Mass radius diagram for transiting planets with CoRoT-3b in the brown
  dwarf desert (\cite{Del08}, Fig.10).}
\label{3-3}       
\end{figure}

\begin{figure}
\begin{center}
\resizebox{1.00\columnwidth}{!}{
  \includegraphics{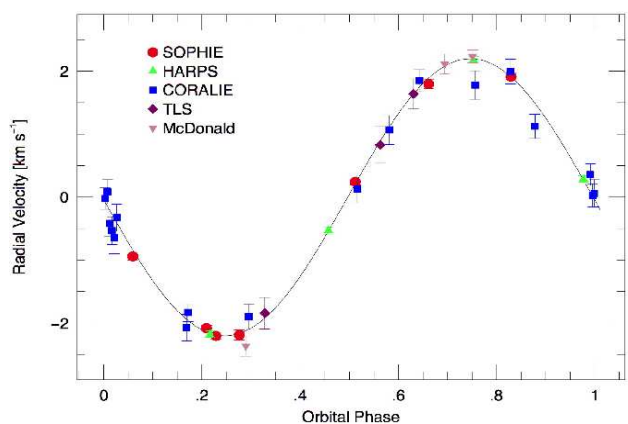} }
\end{center}
\caption{RV measurements of CoRoT-3b (\cite{Del08}, Fig.4.)}
\label{3-4}       
\end{figure}

During the observations of 152 days  from space a series of 34 transits
(Fig. \ref{3-2}) could be measured for a young F3V star with a radius of $1.56 M_{Sun}$ with a
period of 4.26 days (Fig.\ref{3-1}). The
results were a big surprise because with a radius comparable to that of
Jupiter the mass could be determined as $21.66 M_{Jup}$ which clearly
distinguishes this planet from the normal close-in planets (Deleuil et al. \cite{Del08}): either this is a
brown dwarf or member of a new class of planets (Fig. \ref{3-3})  . The determination of the mass
was only possible with RV observations from the ground; in
addition to the four mentioned instruments (HARPS, SOPHIE, CORALIE and TLS,
Fig. \ref{3-4}) and measurements from
McDonald observatory.

\section{CoRoT-4b: a transiting planet in a synchronous orbit}

\begin{figure}
\begin{center}
\resizebox{1.00\columnwidth}{!}{
  \includegraphics{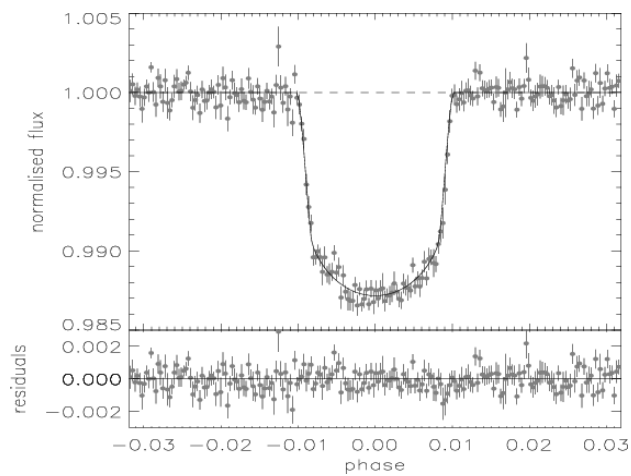} }
\end{center}
\caption{Folded light curve for CoRoT-4b (\cite{Aig08}, Fig.3).}
\label{4-1}       
\end{figure}

\begin{figure}
\begin{center}
\resizebox{1.00\columnwidth}{!}{
  \includegraphics{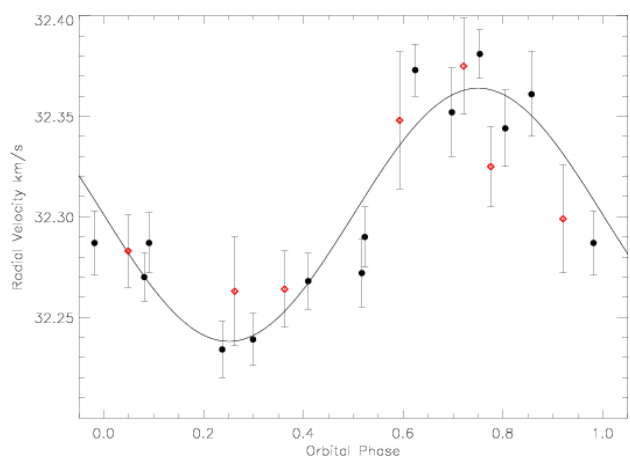} }
\end{center}
\caption{RV measurements of CoRoT-4b with HARPS and SOPHIE (\cite{Mou08}, Fig.2).}
\label{4-2}       
\end{figure}

The planet itself with a radius of 1.2 $R_{Jup}$ and a mass of 0.71 $M_{Jup}$
is orbiting the host star with a period of 9.2 days on a circular orbit.
The planet with its mean density of $0.525 \; g \; cm^{-3}$ is in the expected
location of a Jupiter-like planet in the mass-radius diagram but because of
its period of almost ten days lies outside of the other hot Jupiters with
significantly smaller periods of $P \le 5$ days (Aigrain et al. \cite{Aig08},
Moutou et al. \cite{Mou08}). The reduction in luminosity
caused by the planets' transit (Fig. \ref{4-1}) is in the
order of 1.5 $\%$. The
subsequently undertaken RV measurements (Fig.\ref{4-2}) with HARPS and
SOPHIE confirmed its
planetary nature.
From the lightcurve (not shown here) showing 6 clear transits during the
continuous observation of almost 60 days one can see that the
host star (a young
F0V star of only about 1 Gyr of age) has an active surface with rapidly
evolving starspots. It was possible to derive a rotation period of
P=8.87 days,  which is within its error bars consistent with the
period of the planet. In addition there could be an outer
synchronized envelope of the star with differential surface rotation.

\section{CoRoT-5b: a hot-Jupiter-type planet}

\begin{figure}
\begin{center}
\resizebox{1.00\columnwidth}{!}{
  \includegraphics{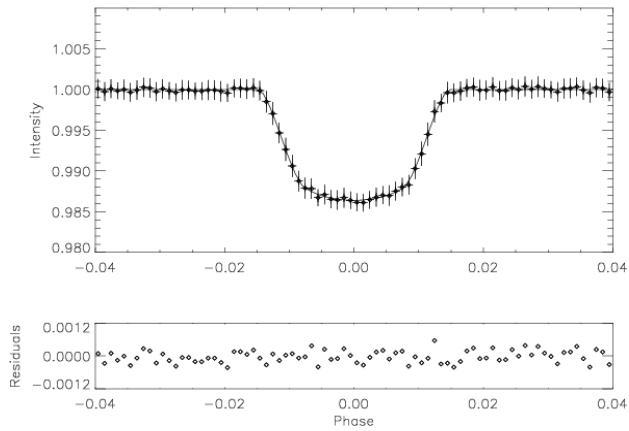} }
\end{center}
\caption{Phase-folded light curve of CoRoT-5b (\cite{Rau09}, Fig.6).}
\label{5-1}       
\end{figure}

\begin{figure}
\begin{center}
\resizebox{1.00\columnwidth}{!}{
  \includegraphics{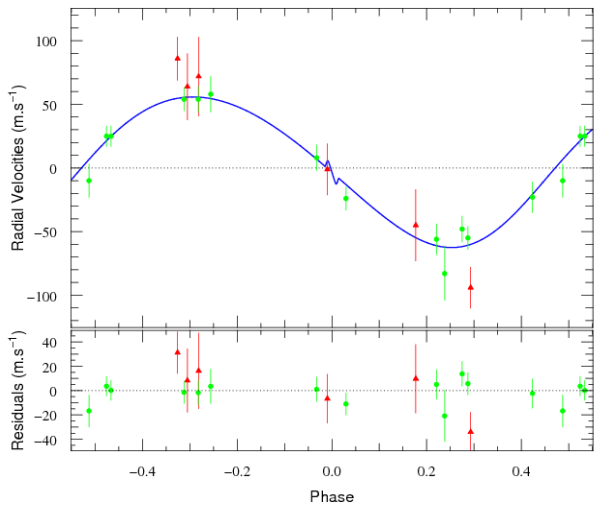} }
\end{center}
\caption{RV measurements of HARPS and SOPHIE showing the Rossiter-Laughlin
  effect of CoRoT-5b during the transit (\cite{Rau09}, Fig.2).}
\label{5-2}       
\end{figure}

After an observation time of 112 days with 27
transits -- Fig. \ref{5-1} shows the light curve with a decrease in flux of
$1.5 \%$ -- the
analysis led to an orbital period of 4 days, a radius of 1.4 $R_{Jup}$
and a mass of  0.47 $M_{Jup}$ (Rauer et al. \cite{Rau09}). The mean density of $0.217 \; g
\; cm^{-3}$
is very low compared to other gas giants observed up to
now\footnote{only the planets WASP-1b and WASP-15b (http://exoplanet.eu) have
  such low densities}. This means that the planet shows the largest radius
anomaly found so far. This relatively old star (5.5-8.3 Gyrs)
is a F9V star with the mass of the sun and a slightly larger radius. Follow-up
observations with SOPHIE and HARPS have been undertaken to derive the RV
curve (Fig.\ref{5-2})
which clearly shows the Rossiter-McLaughlin effect. With additional HARPS
observations it was possible to determine the very low metallicity
(ratio [M/H] =
$-0.25$).

\section{CoRoT-6b: a transiting hot Jupiter planet in a 8.9d orbit around a
low-metallicity star}

\begin{figure}
\begin{center}
\resizebox{1.00\columnwidth}{!}{
  \includegraphics{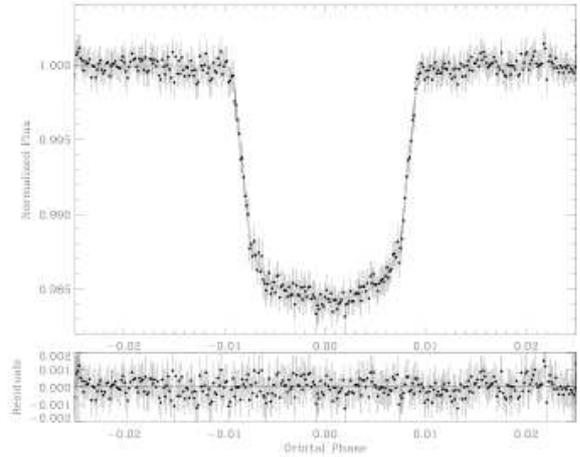} }
\end{center}
\caption{Phase-folded light curve of CoRoT-6b (\cite{Fri09}, Fig.4).}
\label{6-1}       
\end{figure}

\begin{figure}
\begin{center}
\resizebox{1.00\columnwidth}{!}{
  \includegraphics{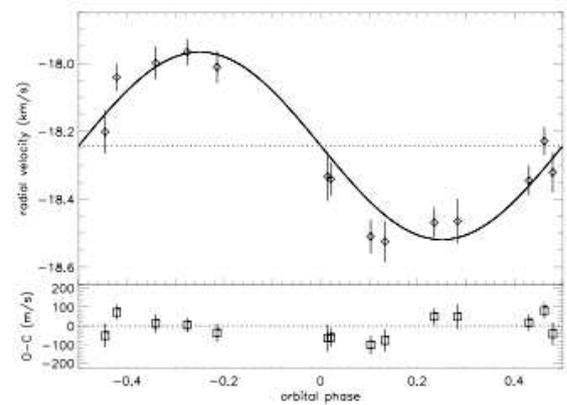} }
\end{center}
\caption{RV measurements of CoRoT-6b with SOPHIE (\cite{Fri09}, Fig.8).}
\label{6-2}       
\end{figure}

This planet discovered by CoRoT turned out to be in a an orbit of P=8.89 with
approximately 3 Jupitermasses and a radius 15 percent larger than
Jupiters' leading to a high density of  $1.94 \; g \; cm^{-3}$ (Fridlund et
al. \cite{Fri09}).
It is hosted by a F9V star with approximately the mass of the Sun and a surprisingly low metallicity
of [$Fe/H= -0.2$]. The combination of the RV measurements with the
photometric ones from space -- respectively
ground-based photometric and spectroscopic observations -- led to the
full characterisation of the planet and the star.
The respective results for the light curve (Fig. \ref{6-1}) and the RV measurements  (Fig. \ref{6-2})
are taken from of (\cite{Fri09}).

\section{CoRoT-7b: the first super-Earth with measured radius}

\begin{figure}
\begin{center}
\resizebox{1.00\columnwidth}{!}{
  \includegraphics{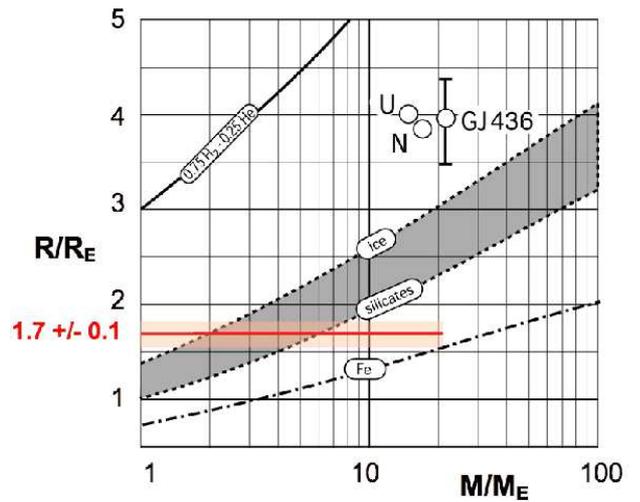} }
\end{center}
\caption{Location of CoRoT7-b in the mass-radius diagram (\cite{Leg09}, Fig.20).}
\label{7-1}       
\end{figure}

The most exciting result of CoRoT up to now is the discovery of
a Super-Earth planet with a terrestrial density
orbiting extremely close to its host star. The G9V star is not older than 2 Gyrs
and has a metallicity of $0.03$ (L\'eger et al. \cite{Leg09}). CoRoT-7b turned
out to be in a circular orbit in a distance of only
approximately 4 times the radius of its host star which could be determined by the analysis of
its 153 transits. The planets' radius is determined
via photometric and spectroscopic measurements and turned out to be slightly smaller than 2 times the
Earth's radius ($r=1.78 \; R_{Earth}$ and
a mass of  $M=4.8 \; M_{Earth}$) \footnote{These values are taken from
  the second paper on this planet (\cite{Que09}); RV measurements in the first paper
  (\cite{Leg09}) were used to exclude non planetary companions.}. The mean
density of $5.6~g~cm^{-3}$ is comparable to the one of the Earth, consequently we can speak of an
Earth-like planet which moves in only 0.854 days around the relatively young
main sequence star which has a rotation period of 23
days. A second planet (Queloz et al. \cite{Que09}) turned out to be
on a 3.89 days orbit and is probably also a Super-Earth
( $m= 8.4 \; M_{Earth}$). It seems that both planets are made of of
rocks or water and rocks similar to our Earth which can be understood from
Fig.\ref{7-1}.

\begin{figure}
\begin{center}
\resizebox{1.00\columnwidth}{!}{
  \includegraphics{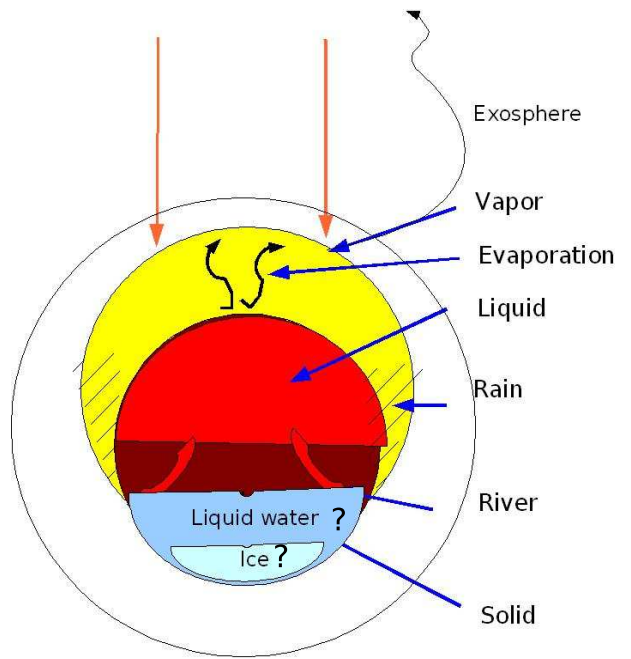} }
\end{center}
\caption{A physical model of CoRoT-7b (\cite{Sch08}).}
\label{jean}       
\end{figure}

Because of the smallest period of CoRoT-7b found for a planet around
another star the temperatures will be extremely high on its surface.
Estimations for the
temperatures on the day side (Schneider \cite{Sch08}) lead to a temperature of
about $2000 K$
depending on the rotation of the planet; due to the vicinity to the star
a bounded rotation is very probable. This would mean that the silicate surface
consists of lava and the planets' atmosphere of evaporated silicates.
On the contrary the far side could be in the temperature range such that
water could be liquid or even be present on its surface in form of ice. In
Fig. \ref{jean} there is a sketch of this planet where bounded rotation because
of the acting tides of
the star is assumed. This -- at least somewhat speculative -- physical model
has been drawn by J. Schneider (\cite{Sch08}) but it characterizes what kind
of completely different appearance we may expect from the 'hells' planet.

\begin{figure}
\begin{center}
\resizebox{1.00\columnwidth}{!}{
  \includegraphics{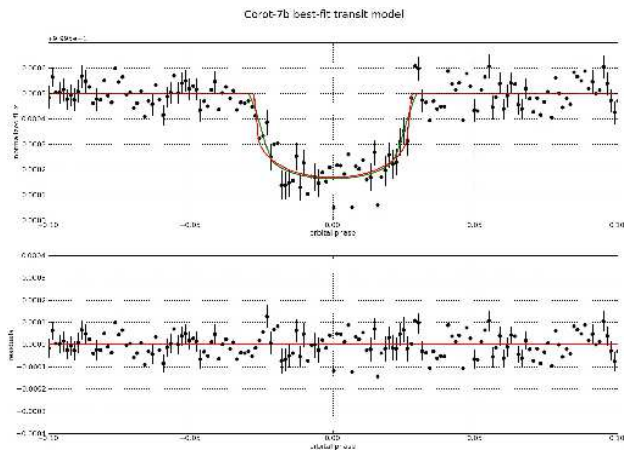} }
\end{center}
\caption{Superposition of 153 transits of CoRoT-7b (\cite{Leg09}, Fig.17).}
\label{7-2}       
\end{figure}

With these measurements for  CoRoT-7b it has been proven that the satellite is
able to discover such planets as
Super-Earth and even smaller ones. From Fig. \ref{7-2}  one can see that the very small amplitude transits of
$\delta F/F   \sim 3.5 \cdot 10^{-4}$ is clearly not the limit of detection.

\section{The Mass loss of the CoRoT planets due to proximity to the host stars}

In a recent study (Lammer et al. \cite{Lam09}) the thermal escape of hydrogen atoms from 57
transiting exoplanets was determined and a modified energy-limited mass loss
equation applied,
which reproduces the full hydrodynamic approach of studies
(e.g. \cite{Pen08}).
By applying the same method to the exoplanets discovered by CoRoT
integrated over the life time of the planetary system we obtain for the given
planetary and stellar parameters the thermal mass loss rates shown in Table \ref{tab:3}.

Depending on the chosen stellar EUV-related heating efficiency $\eta$ of 10 \%
or 25 \% which corresponds to the ratio of the net local gas heating rate to
the rate of  stellar radiative energy absorption, one can see that CoRoT-1b
and CoRoT-5b  lost during their life-time about 1.3 \% and 1.56 \% ($\eta$ = 10 \%)
or 3.07 \% and 3.75 \% ($\eta$ = 25 \%) of their initial masses. The thermal mass
loss of CoRoT-2b, CoRoT-3b, CoRoT-4b and CoRoT-6b are lower and therefore
negligible. The stronger mass loss effect for CoRoT-1b and CoRoT-5b are
related to the lower planetary density, orbital distance and stellar
parameters so that the Roche lobe becomes a relevant factor in the mass loss
evolution
of these planets.

Recently Leitzinger et al. (\cite{Lei09}) presented thermal mass loss
calculations over
evolutionary time scales to investigate whether CoRoT-7b could be a
remnant
of an initially more massive hydrogen-rich gas giant.
The thermal mass loss results of these authors indicate that CoRoT-7b
cannot be a remnant
of a thermally evaporated Jupiter like gas giant. A scenario that
CoRoT-7b is a remnant of an evaporated low
density 'hot sub-Uranus' or 'hot sub-Neptune' is also very unlikely but
can not be completely
excluded. Such an evaporation scenario would need, besides low initial
density of the
planet, also a heating efficiency of about $25 \%$ instead of more realistic
$10 \%$ and an
unlikely short exosphere formation time of less than 50 Myr. From these
simulations
one can conclude that CoRoT-7b was most like\-ly never a gas or ice giant
which has lost
its entire hydrogen envelope, but started its evolution as a
'super-Earth' which
could have lost a thin hydrogen envelope.

The efficiency of the CoRoT exoplanet mass loss is connected to the Roche lobe
effect together with low planetary densities. Depending on the stellar
luminosity spectral type, initial planetary density, stellar EUV-related
heating efficiency, orbital distance, and the connected Roche lobe effect, one
can expect that at orbital distances $\le 0.015$ AU, low density hot gas
giants in orbits around
solar type stars may even evaporate down to their core size.
Therefore, various size and mass type objects should be discovered in the near future.

\section{Conclusions and synopsis of the results}

The observations from space with the satellite CoRoT tur\-ned out to have very
interesting results for the knowledge of the diversity of planets and planetary systems. The
continuous surveillance of 100000 stars for a period of 150 days is very
efficient to detect close by planets (with periods between Venus and Mercury
and smaller) and even allows to 'see' small planets transiting in front of the
star's disk with a very small decrease in the flux like for CoRoT-7b.
But to get absolute planetary parameters
for the mass and the radius it is necessary to make extensive spectroscopic
measurements and also to have a good knowledge of the astrophysical parameters
of the host star. Only after careful analysis of the light curves
(contamination may cause a 'false' alarm) in
combination with thorough ground observation a confirmation of a planet
found around another star can be attested.

The planets found by CoRoT
are of extraordinary value for the extrasolar planetary research in astronomy
because of the complete characterisation of the planets.
One can see that most of the CoRoT planets are giant gas planets (radii
between $0.97 \le R_{Jup} \le 1.49$. The largest ones (CoRoT-1b and 4b)
are also very close to the host star having the shortest periods (1.51 and
1.74 days respectively) which is a clear sign of the important role of the
star radiation on the radius evolution. CoRoT-4 and CoRoT-6 host planets which
have a synchronized period with respect to the host stars' rotation. Concerning
the metallicity it is interesting to note that three of the five hot Jupiters
detected are orbiting metal poor stars, which is in contradiction to the
normal metallicity relations found by RV searches.
As last point it should be emphasized that the domain of planet masses
covered by the CoRoT planets is within a wide range, namely
from the very massive CoRoT-3b planet (with 21.6 $M_{Jup}$) to
the telluric planet CoRoT-7b (0.015 $M_{Jup}$). Another new aspect is the
discovery of close-in planets (CoRoT-7b and 7c). This can be regarded as hint
that --like in our Solar System -- planetary systems are 'full'. This means
that in between two planets no other bigger objects (exceptions are asteroids)
may exist on stable orbits.

The presentation of the major results in this overview is by far not complete -- e.g. we did
not speak about time transit variations (TTV) -- which have also been detected
in some of the data and may be caused by additional planets in the system.
Only the results of the confirmed planets
have been presented\footnote{In the appendix we show the star parameters and
  the derived parameters for the planets as published in the original papers},
but the signals of  many more planets seem to be still in our
data like the ones of planets around double stars. But they can be confirmed
only after difficult and long analysis combined with observations from the ground.

\section{Appendix}

In table 1 the astrophysical parameters of the host stars are listed, in table
2 we listed the orbital parameters of the detected planets and in table 3
the physical parameters are given together with an estimation of the massloss
due to the proximity to the star. All data are taken from the respective
publications given in the text.

\begin{table}
\caption{The astrophysical parameters of the host stars}

\label{tab:1}       
\begin{tabular}{lllll}
\hline\noalign{\smallskip}
star    & type & $M/M_{sun}$ & age [Gyrs] & [Fe/H] \\
\noalign{\smallskip}\hline\noalign{\smallskip}

CoRoT-1 & G0V  & 0.95        & ?          & -0.3 \\
        &      & $\pm 0.15$  &            & $\pm 0.25$ \\

CoRoT-2 & K0V  & 0.97        & ?          & 0.0 \\
        &      & $\pm 0.06$  &            & $\pm 0.1$ \\

CoRoT-3 & F3V  & 1.37        & $2.0_{-0.4}^{+0.8}$ & -0.02 \\
        &      & $\pm 0.09$  &            & $\pm 0.06$ \\

CoRoT-4 & F0V  & $1.1_{-0.02}^{+0.03}$ & $1.0_{-0.3}^{+1.0}$ & 0.0 \\
        &      &             &            & $\pm 0.15$ \\

CoRoT-5 & F9V  & 1.0         & 6.9        & -0.25 \\
        &      & $\pm 0.02$  & $\pm 1.4$  & $\pm 0.06$ \\

CoRoT-6 & F5V  & 1.055       & ?          & -0.2 \\
        &      & $\pm 0.055$ &            & $\pm 0.1$ \\

CoRoT-7 & G9V  & 0.93        & $1.5_{-0.3}^{+0.8}$ & 0.03 \\
        &      & $\pm 0.03$  &            & $\pm 0.06$ \\

\noalign{\smallskip}\hline
\end{tabular}
\end{table}

\begin{table}
\caption{The orbital elements of the planets}
\label{tab:2}       
\begin{tabular}{lllll}
\hline\noalign{\smallskip}
planet & Period [d]   & a [AU]                   & inc [deg]          &
ecc \\
\noalign{\smallskip}\hline\noalign{\smallskip}

CoRoT-1b & 1.508956     & 0.0254                   & 85.1               & 0 \\
       & $\pm 6e-6$   & $\pm 0.0004$             & $\pm 0.5$          & \\
\noalign{\smallskip}

CoRoT-2b & 1.742996     & 0.0281                   & 87.84              & 0 \\
       & $\pm 2e-6$   & $\pm 0.0009$             & $\pm 0.1$          & \\
\noalign{\smallskip}

CoRoT-3b & 4.2568       & 0.057                    & 85.9               & 0 \\
       & $\pm 5e-6$   & $\pm 0.003$              & $\pm 0.8$          & \\
\noalign{\smallskip}

CoRoT-4b & 9.20205      & 0.090                    & 90.0               &
$0_{-0}^{+0.1}$ \\
       & $\pm 37e-5$  & $\pm 0.001$              & $_{-0.085}^{+0}$   & \\
\noalign{\smallskip}

CoRoT-5b & 4.037896     & 0.04947                  &  85.83             &
0.09 \\
       & $\pm 2e-6$   & $_{-0.00029}^{+0.00026}$ & $_{-1.38}^{+0.99}$ &
$_{-0.04}^{+0.09}$ \\
\noalign{\smallskip}

CoRoT-6b & 8.886593     & 0.0855                   & 89.07              & $<
0.1$ \\
       & $\pm 4e-6$   & $\pm 0.0015$             & $\pm 0.3$          & \\
\noalign{\smallskip}

CoRoT-7b & 0.85358      & 0.0172                   & 80.1               & 0 \\
       & $\pm 2e-5$   & $\pm 0.0003$             & $\pm 0.3$          & \\
\noalign{\smallskip}

CoRoT-7c & 3.698        & 0.046                    & ?                  & 0 \\
       & $\pm 0.003$  &                          &                    & \\

\noalign{\smallskip}\hline
\end{tabular}
\end{table}

\begin{table}
\caption{The physical properties of the planets}
\label{tab:3}
\begin{tabular}{lllll}
\hline\noalign{\smallskip}
planet & $M/M_{jup}$               & $R/R_{jup}$               & $\varrho
[g\;cm^{-3}]$ & Massloss [$\%$]\\
\noalign{\smallskip}\hline\noalign{\smallskip}

CoRoT-1b & $1.03$          & $1.49$           & 0.38 & 1.3--10.81\\
 &  $\pm 0.12$ & $ \pm 0.08$ & & \\
\noalign{\smallskip}

CoRoT-2b & $3.31$           & $1.465$          & 1.31 & 0.06--0.59\\
 & $\pm 0.16$ &  $\pm 0.029$ & & \\

\noalign{\smallskip}

CoRoT-3b & $21.66$           & $1.01$           & 26.4 & 0.0002--0.0017\\
 & $\pm 1.00$ & $\pm 0.07$ & & \\
\noalign{\smallskip}

CoRoT-4b & $0.72$          & $1.19$    & 0.525 & 0.12--1.212\\
 &  $\pm 0.08$  &  $_{-0.05}^{+0.06}$ & & \\
\noalign{\smallskip}

CoRoT-5b & $0.467$ & $1.388$  & 0.217 & 1.56-- 12.78\\
 & $ _{-0.024}^{+0.067}$ & $_{-0.047}^{+0.046}$ & & \\
\noalign{\smallskip}

CoRoT-6b & $2.96$           & $1.166$         & 1.94 & $\le$ 1\% \\
 &  $\pm 0.34$ &  $\pm 0.035$ & & \\
\noalign{\smallskip}

CoRoT-7b & $0.015$          & $0.172$      & 4.23 & ?\\
 & $\pm 0.003$ & $_{-0}^{+0.022}$ & & \\
\noalign{\smallskip}

CoRoT-7c & $0.026$          & ?                         & ? & ?\\
 &  $\pm 0.003$ & & & \\
\end{tabular}
\end{table}

\section{Acknowledgements}

The LAM team and the Austrian team thanks CNES respectively ASA for funding the CoRoT project. The IAC team acknowledges the support of grants of the Education and Science Ministry. The German team (TLS and Univ. Cologne) acknowledges the support of DLR grants. For reading the manuscript and preparing an appropriate outlay of the pictures R.~D. needs to thank A.~Bazs\'o and V.~Eybl.

\end{document}